%% file: Template.tex
\documentclass{article}
\usepackage{spconf,amsmath,graphicx,hyperref,comment,booktabs,multicol,siunitx,amssymb,multirow}
\usepackage{booktabs}
\usepackage{bm}
\usepackage{enumitem}
\setlist[itemize]{nosep} 
\setlist[enumerate]{nosep} 

\title{Geneses: Unified Generative Speech Enhancement and Separation}
\name{Kohei Asai$^1$, Wataru Nakata$^1$, Yuki Saito$^{1,2}$, Hiroshi Saruwatari$^1$}
\address{$^1$The University of Tokyo, Japan,\\$^2$ National Institute of Advanced Industrial Science and Technology (AIST), Japan.}
\begin{document}
    \ninept
\setlength\intextsep{10pt} 
\setlength\textfloatsep{10pt} 
\setlength{\dbltextfloatsep}{10pt} 
\setlength{\dblfloatsep}{10pt}

\maketitle

\begin{abstract}
Real-world audio recordings often contain multiple speakers and various degradations, which limit both the quantity and quality of speech data available for building state-of-the-art speech processing models.
Although end-to-end approaches that concatenate speech enhancement (SE) and speech separation (SS) to obtain a clean speech signal for each speaker are promising, conventional SE-SS methods suffer from complex degradations beyond additive noise.
To this end, we propose \textbf{Geneses}, a generative framework to achieve unified, high-quality SE--SS.
Our Geneses leverages latent flow matching to estimate each speaker's clean speech features using multi-modal diffusion Transformer conditioned on self-supervised learning representation from noisy mixture.
We conduct experimental evaluation using two-speaker mixtures from LibriTTS-R under two conditions: additive-noise-only and complex degradations.
The results demonstrate that Geneses significantly outperforms a conventional mask-based SE--SS method across various objective metrics with high robustness against complex degradations.
Audio samples are available in our demo page\footnote{\url{https://freckle-heron-b84.notion.site/Geneses-Sample-295b7fa15f5580adbb2fefa0ad5dd348}}.
\end{abstract}

\input{sections/introduction}

\input{sections/related_work}

\input{sections/method}

\input{sections/experiment}

\input{sections/conclusion}


\bibliographystyle{IEEEtran}
\bibliography{refs}

\end{document}

%% file: sections/introduction.tex
\section{Introduction}

Speech separation (SS)~\cite{wang2018supervised}, which aims to isolate individual speaker signals from a mixture of multiple voices, is a key technology for robust communication through speech processing technologies such as automatic speech recognition.
In practical scenarios, recorded speech is often degraded not only by overlapping speakers but also by various real-world factors such as additive noise, reverberation, codec distortion, and packet loss.
To make such degraded speech high-quality and intelligible, speech enhancement (SE)~\cite{nuthakki21se}---which converts a clean signal from degraded one---should be addressed together with SS.
A unified framework that jointly performs SE and SS can provide consistently high performance across diverse acoustic environments, improving the usability of speech-based telepresence and extended reality applications.

Hu et al.~\cite{hu2023unifying} proposed a unified approach for noise-robust SS by combining SE and SS as a single deep neural network (DNN)-based model.
Their approach adopts a cascaded SE--SS architecture optimized via multi-task learning and a gradient modulation (GM) strategy, which suppresses conflicts between SE and SS to mitigate the over-suppression of speech components.
Although this approach showed promising results under ``additive-noise-only'' conditions, it cannot handle complex degradations due to the limitation of mask-based SE--SS, restricting its applicability in more realistic environments.

To this end, we propose \textbf{Geneses}\footnote{\textbf{Gene}rative \textbf{s}peech \textbf{e}nhancement and \textbf{s}eparation}, a unified generative approach that jointly performs SE and SS.
Our Geneses leverages flow matching~\cite{lipman2023flow} on latent feature representations to estimate each speaker's clean speech features using multi-modal diffusion Transformer (MM-DiT)~\cite{esser2024scaling} conditioned on self-supervised learning (SSL) representation from noisy mixture.
We conduct experimental evaluation using two-speaker mixtures from LibriTTS-R~\cite{koizumi2023libritts} under two conditions: additive-noise-only and complex degradations.
The results demonstrate that Geneses significantly outperforms a conventional Hu et al.'s noise-robust SS method~\cite{hu2023unifying} across various objective metrics with high robustness against complex degradations.

%% file: sections/related_work.tex
\section{Related Work}

\subsection{DNN-based SS and Noise-robust SS}

SS has primarily evolved as a task to separate individual speaker signals from multi-talker mixtures in clean environments. 
Developments on sophisticated DNN architectures such as Transformers (e.g., SepFormer~\cite{subakanattention}) and generative modeling like flow matching (e.g., FLOSS~\cite{scheibler2025source}) have achieved high-quality separation.
However, these SS models are designed under the premise of that the mixture consists of clean speech. 

To address this issue, noisy SS, which aims to perform noise removal and speaker separation simultaneously using a single model, has been proposed.
For example, noise-aware SS~\cite{zhang2024noise} explicitly treats noise as an additional source to suppress its leakage into the separated speech, and GeCo~\cite{wang2024noise} post-processes the output of a discriminative model using a generative model to remove noise and unnatural distortions. 
Although these attempts enlarges the application range of SS, conventional approaches only considered additive noise as the degradation case.
Therefore, noise-robust SS under complex degradations in more realistic scenarios, such as clipping, codec artifacts, and packet loss, remains a significant challenge.

\subsection{Universal SE for Complex Degradations}

In the field of SE, universal SE, which aims to restore clean speech from complex degradations using a single model, has been actively studied. 
Especially, universal SE incorporating generative models, particularly score-based diffusion models~\cite{serrauniversal}~\cite{scheibleruniversal}, have demonstrated promising results.
The reason is that they can generate high-quality speech through the stochastic process from degraded input to clean signal, which is not uniquely determined due to various degradations, such as bandwidth extension and packet loss.

For example, SGMSE+~\cite{richter2023speech} adapts the diffusion process using a stochastic differential equation to directly model the transformation from clean to corrupted speech (including noise and reverberation). 
Another notable approach is UNIVERSE++~\cite{scheibleruniversal}, which improves upon the UNIVERSE model~\cite{serrauniversal} by incorporating architectural improvements, adversarial training to promote high-quality feature extraction, and low-rank adaptation (LoRA)~\cite{hulora} to preserve linguistic content. 
Although these methods demonstrated the potential of diffusion models to tackle a diverse set of degradations effectively, they have predominantly focused on enhancing single-speaker speech. 
That is, they did not address the challenge of separating speech when multiple speakers are present simultaneously within the complexly degraded signal.

By leveraging insights from universal SE, this research aims to enable SS to operate effectively even in environments containing complex degradations.

%% file: sections/method.tex
\section{Geneses}

\begin{figure}[tb]
  \centering
  \includegraphics[width=\linewidth]{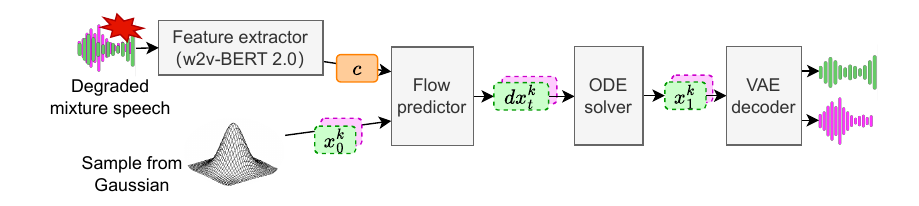}
  \caption{Inference pipeline of Geneses. Flow matching is performed on the latent representations of a pre-trained VAE.}
  \label{fig:geneses_inference}
\end{figure}

\begin{figure}[tb]
  \centering
  \includegraphics[width=\linewidth]{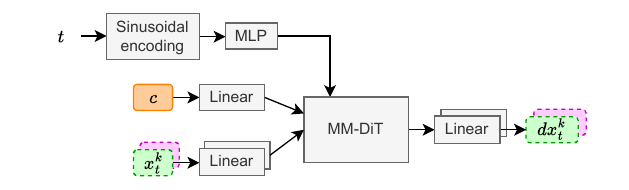}
  \caption{Architecture of flow predictor.}
  \label{fig:geneses}
\end{figure}

Geneses proposed in this study aims to solve SS and SE in a unified manner.
Based on the recent success of flow matching in universal SE and generative SS, Genesis also uses flow matching for this task. 
Specifically, we employ latent flow matching~\cite{dao2023flow} similar to previous work~\cite{guimaraes2025ditse} because learning to generate raw waveform is difficult.
Figure~\ref{fig:geneses_inference} shows the overview of inference by Geneses consisting of three modules: input feature extractor, variational auto encoder (VAE)~\cite{kingma2014auto}, and flow predictor.
The extractor leverages features from the input degraded mixture signal to provide sufficient information to make the predictor well estimate the flow towards VAE latent features of clean speech signals.

In this paper, we mainly focus on SE--SS with a two-speaker mixture case. However, we expect that Geneses can be extended to more challenging cases like three or four speakers.

\subsection{Input Feature Extractor}
In order to extract meaningful information from the highly degraded speaker-mixed speech signal, using a robust feature extractor is important. Therefore, we use a pre-trained SSL model, w2v-BERT 2.0~\cite{barrault2023seamless}.
w2v-BERT 2.0\footnote{\url{https://huggingface.co/facebook/w2v-bert-2.0}} is trained on 4.5M hours of speech spanning across 143 languages, making it highly suited for extracting robust features from degraded signals as shown in previous work~\cite{nakata2025sidon}. 
Furthermore, given this large-scale training, it is also expected to get rich information robustly from mixed speech.
The extracted features $\bm{c}$ are processed through a linear layer before being fed into the Transformer-based flow predictor described later. 
In line with previous studies~\cite{nakata2025sidon}, we also finetune this w2v-BERT model using LoRA~\cite{hulora} in order to prevent catastrophic forgetting of the rich domain knowledge acquired through SSL.

\subsection{VAE for Latent Flow Matching and Speech Synthesis}
Instead of directly handling the raw audio waveform, a VAE is used to compress the audio into a more manageable lower-dimensional latent space, where the latent flow matching process occurs.
Let $t \in [0, 1]$ denote the timestep variable for the flow-matching process.
Here, $\bm{x}_0$ is defined as a sample from a standard Gaussian distribution, and $\bm{x}_1$ is the VAE representation of the target clean speech. 
Since we use rectified flows~\cite{liu2023flow}, $\bm{x}_t$ is expressed as $\bm{x}_t=(1-t)\bm{x}_0+t\bm{x}_1$.
For the unified SE--SS task, this latent space representation is divided into two independent tracks, $\bm{x}_t^1$ and $\bm{x}_t^2$, corresponding to the target speakers. 
These latent representations define the space where the flow matching process occurs. 
Each latent representation $\bm{x}_t^k\,(k\in\{1,2\})$ is passed through a dedicated linear layer to match the dimensionality to the flow predictor. 
For the VAE architecture, we adopt a model based on the Descript Audio Codec (DAC)~\cite{kumar2023high}, similar to previous work ~\cite{guimaraes2025ditse}.

In the following sections, we denote the combined latent representation for both speakers at the timestep $t$ as $\bm{x}_t$, which typically represents the stacking of the individual latent vectors $[\bm{x}_t^1,\,\bm{x}_t^2]$. The same notation applies to $\bm{x}_0$ and $\bm{x}_1$.

\subsection{Flow Predictor}

The flow predictor parametrized by $\theta$, $\bm{v}_{\theta}(\bm{x}_t, \bm{c}, t)$ is the core of our Geneses, which predicts this conditional vector field. 
It takes the current latent representation $\bm{x}_t$ (a state on the trajectory between the noise $\bm{x}_0$ and the target $\bm{x}_1$), the SSL conditional features $\bm{c}$, and timestep $t$ as input, and outputs the predicted a vector towards the clean, separated target representation $\bm{x}_1$. 

We employ the MM-DiT~\cite{esser2024scaling} architecture for this predictor.
MM-DiT utilizes separate weights for modalities but joins their sequences during the attention operation, which enables a bidirectional flow of information between them.
It can effectively fuse different features from the VAE latent representations $\bm{x}_t$, SSL model features $\bm{c}$, and timestep information $t$---through its attention mechanism.
The timestep $t$ is encoded using standard sinusoidal encoding~\cite{sunlearning} followed by a multi-layer perceptron (MLP), and is supplied as a conditioning vector to all layers of the MM-DiT.
The MM-DiT takes the linearly processed $\bm{x}_t, \bm{c}$ and the MLP-transformed $t$ as input, and outputs the predicted flow vector field $\bm{v}_{\theta}$.

During inference, we sample $\bm{x}_0\sim\mathcal{N}(0,I)$ and solve the ODE (Eq. \ref{ode}) numerically using the predicted vector field to obtain $\bm{x}_1$, which is then decoded by the VAE:
\begin{equation}
    \frac{d\bm{x}_t}{dt} = \bm{v}_{\theta}(\bm{x}_t, \bm{c}, t). \label{ode}
\end{equation}

\subsection{Training}

The model parameters of Geneses $\theta$ are optimized to minimize the mean squared error (MSE) loss between the predicted vector field and the target vector field.
This loss function is expressed as:
\begin{equation}
    \mathcal{L}(\theta)=\mathrm{MSE}(\bm{v}_{\theta}(\bm{x}_t,\bm{c},t),\,\bm{x}_1-\bm{x}_0).
\end{equation}
During training, two components are updated: the flow predictor and the LoRA parameters applied to the pre-trained w2v-BERT 2.0 feature extractor.
Conversely, the parameters of the pre-trained VAE are frozen. 

Note that unlike conventional discriminative separation methods, we do not employ Permutation Invariant Training (PIT) loss.
Since our objective is to regress the vector field pointing to a specific target instance $\bm{x}_1$ (the stacked latent representations), applying PIT would introduce ambiguity into the target direction, hindering the learning of a consistent flow.



%% file: sections/experiment.tex
\section{Experiment}

\begin{table*}[h!]
  \centering
  \caption{Reference-free evaluation results. Better results among comparative methods are in \textbf{bold}. Note that the UTMOSv2 score can be less than 1, which is an inherent characteristic of the model.}
    \label{tab:results_noise_ref_free}
  \small 
  \begin{tabular}{lrrrrrrrr}
    \toprule
    & \multicolumn{4}{c}{\textbf{Background Noise Only}} & \multicolumn{4}{c}{\textbf{Complex Degradations}} \\
    \cmidrule(lr){2-5} \cmidrule(lr){6-9}
    \textbf{Method} & \textbf{DNSMOS} $\uparrow$ & \textbf{NISQA} $\uparrow$ & \textbf{UTMOSv2} $\uparrow$ & \textbf{WER} $\downarrow$ & \textbf{DNSMOS} $\uparrow$ & \textbf{NISQA} $\uparrow$ & \textbf{UTMOSv2} $\uparrow$ & \textbf{WER} $\downarrow$ \\
    \midrule
    Ground Truth & 3.37 & 4.73 & 3.65 & 0.11 & 3.37 & 4.73 & 3.65 & 0.11 \\
    Conventional & 2.91 & 2.32 & 1.75 & \textbf{0.35} & 2.08 & 1.34 & 0.84 & 5.54 \\
    Geneses (proposed) & \textbf{3.40} & \textbf{4.44} & \textbf{3.44} & 0.39 & \textbf{3.39} & \textbf{4.44} & \textbf{3.40} & \textbf{0.43} \\
    \bottomrule
  \end{tabular}
\end{table*}

\begin{table*}[h!]
  \centering
  \caption{Reference-aware evaluation results. Better results among comparative methods are in \textbf{bold}.}
   \label{tab:results_noise_ref_aware}
  \small 
  \begin{tabular}{lrrrrrrrrrr}
    \toprule
    & \multicolumn{5}{c}{\textbf{Background Noise Only}} & \multicolumn{5}{c}{\textbf{Complex Degradations}} \\
    \cmidrule(lr){2-6} \cmidrule(lr){7-11}
    \textbf{Method} & \textbf{ESTOI} $\uparrow$ & \textbf{MCD} $\downarrow$ & \textbf{LSD} $\downarrow$ & \textbf{SBS} $\uparrow$ & \textbf{SpkSim} $\uparrow$ & \textbf{ESTOI} $\uparrow$ & \textbf{MCD} $\downarrow$ & \textbf{LSD} $\downarrow$ & \textbf{SBS} $\uparrow$ & \textbf{SpkSim} $\uparrow$ \\
    \midrule
    Conventional & 0.72 & \textbf{7.17} & \textbf{3.96} & 0.80 & 0.95 & 0.42 & 9.24 & 5.79 & 0.61 & 0.91 \\
    Geneses (proposed) & \textbf{0.75} & 7.60 & 4.65 & \textbf{0.83} & \textbf{0.99} & \textbf{0.72} & \textbf{8.09} & \textbf{5.00} & \textbf{0.82} & \textbf{0.98} \\
    \bottomrule
  \end{tabular}
\end{table*}

\subsection{Experimental Condition}

\subsubsection{Data Preparation}

We used LibriTTS-R~\cite{koizumi2023libritts}, which is an English audiobook speech containing 585 hours by 2,456 speakers.
Mixtures were created by taking a sum of two utterances from different speakers.
The splits for the training, validation, and test sets followed the official subsets provided by the LibriTTS-R corpus.
The numbers of mixed speech samples were 110,000 for the training set, 5,000 for the validation set, and 1,500 for the test set.
We then prepared two types of degraded audio from these mixtures as follows.

\noindent\textbf{Data with Background Noise Only:}
Background noise was selected from the DNS5 Challenge~\cite{dubey2024icassp}, WHAM!~\cite{wichern2019wham}, FSD50K~\cite{fonseca2021fsd50k}, the Free Music Archive, and wind noise simulated by the SC-Wind-Noise-Generator~\cite{mirabilii2022simulating}.
This noise was added to the clean speech at a signal-to-noise ratio (SNR) randomly chosen from a uniform distribution $\mathcal{U}(-5, 20)\,\si{dB}$.

\noindent\textbf{Data with Complex Degradations:}
The degradation process followed as: 1) reverberation, 2) noise addition, 3) bandwidth limitation, 4) clipping, 5) codec degradation, and 5) packet loss, each with a 50\% probability of being applied. See \cite{nakata2025sidon} for the implementation details. 
Note that, for the test set, unlike previous work~\cite{nakata2025sidon}, we used the DEMAND dataset\footnote{\url{https://www.kaggle.com/datasets/chrisfilo/demand}} as the noise dataset and the MIT Environmental Impulse Response Dataset as the RIR dataset\footnote{\url{https://huggingface.co/datasets/davidscripka/MIT_environmental_impulse_responses}}.

\subsubsection{DNN Architecture}

\noindent\textbf{Feature Extractor:} We utilized w2v-BERT 2.0 as the foundation, extracting 1,024-dimensional features from 16~kHz audio. The model was fine-tuned by LoRA with rank 64, targeting the output dense layers with 8 hidden adapters ($\alpha=16,\,\mathrm{dropout}=0.1$).

\noindent\textbf{VAE:}
We pretrained the VAE using train subsets of LibriTTS-R. The VAE operated at 24~kHz.
The encoder rates were set to $2,3,4,5,8$ resulting in the 25~Hz frame rate of latent features.
The latent dimension of the VAE were set to 16.
For other settings, we followed the previous work~\cite{kumar2023high}.

\noindent\textbf{Flow Predictor:} The MM-DiT consisted of 12 Transformer layers with 768 hidden dimensions and 12 attention heads. It processed concatenated SSL features (1,024-dim) and VAE latents (16-dim per speaker) using FlashAttention~\cite{dao2022flashattention} with RMSNorm~\cite{jiang2023pre}. To encode the temporal position within the feature sequences, sinusoidal positional embeddings supported sequences up to 20 seconds of the input signal. 
This positional encoding represented the sequential order of the features, distinct from the flow matching timestep $t$.
Time conditioning was achieved through a 256-dimensional timestep embedder followed by a two-layer MLP with SiLU activation.

\subsubsection{Optimization and Inference}

For sampling the timestep $t$ during training, we used logit normal sampling~\cite{esser2024scaling}.
For optimization, we employed the AdamW~\cite{loshchilovdecoupled} optimizer with $\beta_1=0.9$, $\beta_2 = 0.999$ and weight decay of $10^{-4}$.
Learning rate scheduling was performed for improving the stability of training.
The learning rate was set to $10^{-5}$.
The model was trained for a total of 150k iterations. 
All experiments were conducted on a system with 8 NVIDIA H200 GPUs, using a total batch size of 16.
During inference, the ODE (Eq. \ref{ode}) was numerically solved using the Euler method with a fixed step size of 0.01.

\subsubsection{Evaluation Criteria}

For the objective evaluation, we used two sets of metrics: reference-free and reference-aware.

The reference-free metrics included DNSMOS~\cite{reddy2021dnsmos}, NISQA~\cite{mittag2021nisqa}, and UTMOSv2~\cite{baba2024t05}, which are deep learning-based models that predict subjective speech quality, and word error rate (WER) obtained from the Whisper ASR system~\cite{radford2023robust}\footnote{\url{https://huggingface.co/openai/whisper-large-v3}}.

The reference-aware metrics consisted of  ESTOI (extended short-time objective intelligibility)~\cite{jensen2016algorithm} to evaluate speech intelligibility, MCD (mel-cepstral distortion)~\cite{kubichek1993mel} to assess timbral distortion, LSD (log-spectral distance)~\cite{gray2003distance} to evaluate overall spectral distortion, SpeechBERTScore (SBS)~\cite{saeki2024speechbertscore}\footnote{The SSL model used to calculate SpeechBERTScore is \url{https://huggingface.co/utter-project/mHuBERT-147}}, which is highly correlated with human subjective ratings, and speaker similarity (SpkSim) calculated with x-vectors~\cite{snyder2018x}\footnote{\url{https://huggingface.co/speechbrain/spkrec-xvect-voxceleb}}.

For comparison, we employed the method proposed by Hu et al.~\cite{hu2023unifying}.
The same dataset as our proposed method was used, and the hyperparameters were set to the exact values reported in their paper.

\subsection{Result}

The evaluation results for the reference-free and reference-aware metrics are presented in Table \ref{tab:results_noise_ref_free} and Table \ref{tab:results_noise_ref_aware}, respectively. 

\subsubsection{Results under Background Noise Only Condition}

Looking at the reference-free metrics, Geneses significantly outperformed the conventional method on the subjective quality prediction models DNSMOS, NISQA, and UTMOSv2, achieving scores close to the Ground Truth. 
However, for WER obtained using Whisper ASR, the conventional method performed slightly better than Geneses.
In the reference-aware metrics, Geneses showed superior results compared to the conventional method for ESTOI, SBS, and SpkSim. 
Conversely, the conventional method achieved slightly better scores for MCD and LSD.
These results indicate that Geneses achieves high perceptual quality while slightly degrades scores on objective metrics such as WER, LSD, and MCD compared to the conventional method. 
This observation reflects a well-known trade-off inherent in generative SE.
As demonstrated by Scheibler et al.~\cite{scheibleruniversal}, generative SE models often struggle with content preservation, exhibiting ``high LSD and WER''.
This is attributed to the models' tendency to ``change content or drop segments,'' a phenomenon often described as hallucinations~\cite{kalai2025language}.
In contrast, purely discriminative models (e.g., BSRNN) typically excel at these fidelity metrics, showing ``low LSD'' and ``low WER,'' but often do so at the expense of ``naturalness.'' 
Therefore, the slight degradation in these objective fidelity scores for our model is a known characteristic of generative approaches that prioritize producing natural-sounding speech over strictly minimizing waveform reconstruction or linguistic errors.

In summary, while there was a slight trade-off in objective fidelity metrics like WER, LSD, and MCD---a known challenge in generative SE approaches---the overall results were highly positive, driven by substantial improvements in perceptual quality. 
This demonstrates that Geneses works robustly and effectively, even within conventional experimental settings.


\subsubsection{Results under Complex Degradations Condition}

In the reference-free metrics, Geneses substantially surpassed the conventional method across all metrics (DNSMOS, NISQA, UTMOSv2), achieving quality close to the Ground Truth shown in Table~\ref{tab:results_noise_ref_aware}. Notably, regarding WER, the conventional method's performance degraded to a level where recognition is difficult (5.54), whereas Geneses significantly reduced WER (0.43).
Similarly, for the reference-aware metrics, Geneses demonstrated considerably better results than the conventional method across all metrics (ESTOI, MCD, LSD, SBS, SpkSim). 
This signifies that Geneses is far superior in terms of intelligibility, spectral distortion, correlation with subjective quality, and speaker similarity.

In summary, these results clearly demonstrate the limitations of the conventional method when faced with speech corrupted by complex, real-world degradations. 
The conventional approach fails to handle such challenging conditions, as evidenced by its failures across all metrics, including a WER (5.54) that renders the speech unrecognizable. 
In contrast, Geneses not only avoids this failure but also robustly enhances the audio, delivering high perceptual quality, intelligibility, and speaker fidelity.
This demonstrates that our proposed generative approach is far superior and effectively capable of separating and restoring speech from severe and multifaceted degradations where traditional methods completely break down.


%% file: sections/conclusion.tex
\section{conclusion}

In this paper, we proposed Geneses, a speech separation and enhancement framework designed to simultaneously address both ``complex degradations''---including clipping and codec distortion, not just background noise---and the mixture of multiple speakers.
Experimental results showed that our proposed method is highly effective against such complex degradations, whereas conventional methods degrade significantly. Future work includes applying this method to non-simulated speaker-mixed speech, such as actual conversational data.

{\bf Acknowledgements:}
This work was supported by JST Moonshot JPMJMS2011, JSPS KAKENHI 25KJ0806, JST BOOST JPMJBY24C9, Research Grant S of the Tateisi Science and Technology Foundation and the AIST policy-based budget project ``R\&D on Generative AI Foundation Models for the Physical Domain.''